\documentclass[aps,prl,twocolumn,showpacs]{revtex4}

\usepackage{graphicx}
\usepackage{amsmath,amscd,amssymb}
\usepackage{color}

\newcommand{\be}{\begin{equation}}
\newcommand{\ee}{\end{equation}}
\newcommand{\bea}{\begin{eqnarray}}
\newcommand{\eea}{\end{eqnarray}}
\newcommand{\br}{\mathbf{r}}

\newcommand{\lan}{\left\langle}
\newcommand{\ran}{\right\rangle}

\begin{document}

\title{Ionic Capillary Evaporation in Weakly Charged Nanopores}

\author{Sahin Buyukdagli, Manoel Manghi, and John Palmeri}
\affiliation{Laboratoire de Physique Th\'eorique  -- IRSAMC, CNRS and Universit\'e de Toulouse, UPS, F-31062 Toulouse, France}
\date{\today}

\begin{abstract}
Using a variational field theory, we show that an electrolyte confined to a neutral cylindrical nanopore {traversing a low dielectric membrane} exhibits a first-order ionic liquid-vapor pseudo phase transition from an ionic-penetration ``liquid'' phase to an ionic-exclusion ``vapor'' phase, controlled by nanopore modified ionic correlations and dielectric repulsion. For weakly charged nanopores, this pseudo transition survives and may shed light on the mechanism behind the rapid switching of nanopore conductivity observed in experiments.
\end{abstract}

\maketitle

Electrolytes near charged surfaces are omnipresent in soft-matter (charged colloidal suspensions, polyelectrolyte mixtures), biology (DNA, proteins and cell membranes)~\cite{review} and nanofiltration (ion channels)~\cite{yarosh,shklovskii}.
For bulk ionic fluids, the existence of an ionic first-order liquid-vapor (L-V) phase transition is now well established~\cite{bulk} with the driving mechanism being the competition between the short range steric repulsion and the attractive correlations arising from the long range Coulomb interaction. However, for aqueous bulk electrolytes composed of conventional inorganic salts, the critical temperature is well below freezing {($T_{\rm c}^{\rm bulk}\approx 50$~K)}, so that it can be reached experimentally only with special liquids~\cite{bulk}.
When electrolytes are in contact with low dielectric, and possibly charged, mesoscopic bodies,  interactions between mobile ions and the body surface come into play and strongly modify ion-ion interactions~\cite{surface,onsager}. {More generally, the influence of confinement on L-V transitions is of broad fundamental and technological interest~\cite{evaporation}.}

\begin{figure}[t]
\includegraphics[width=0.8\linewidth]{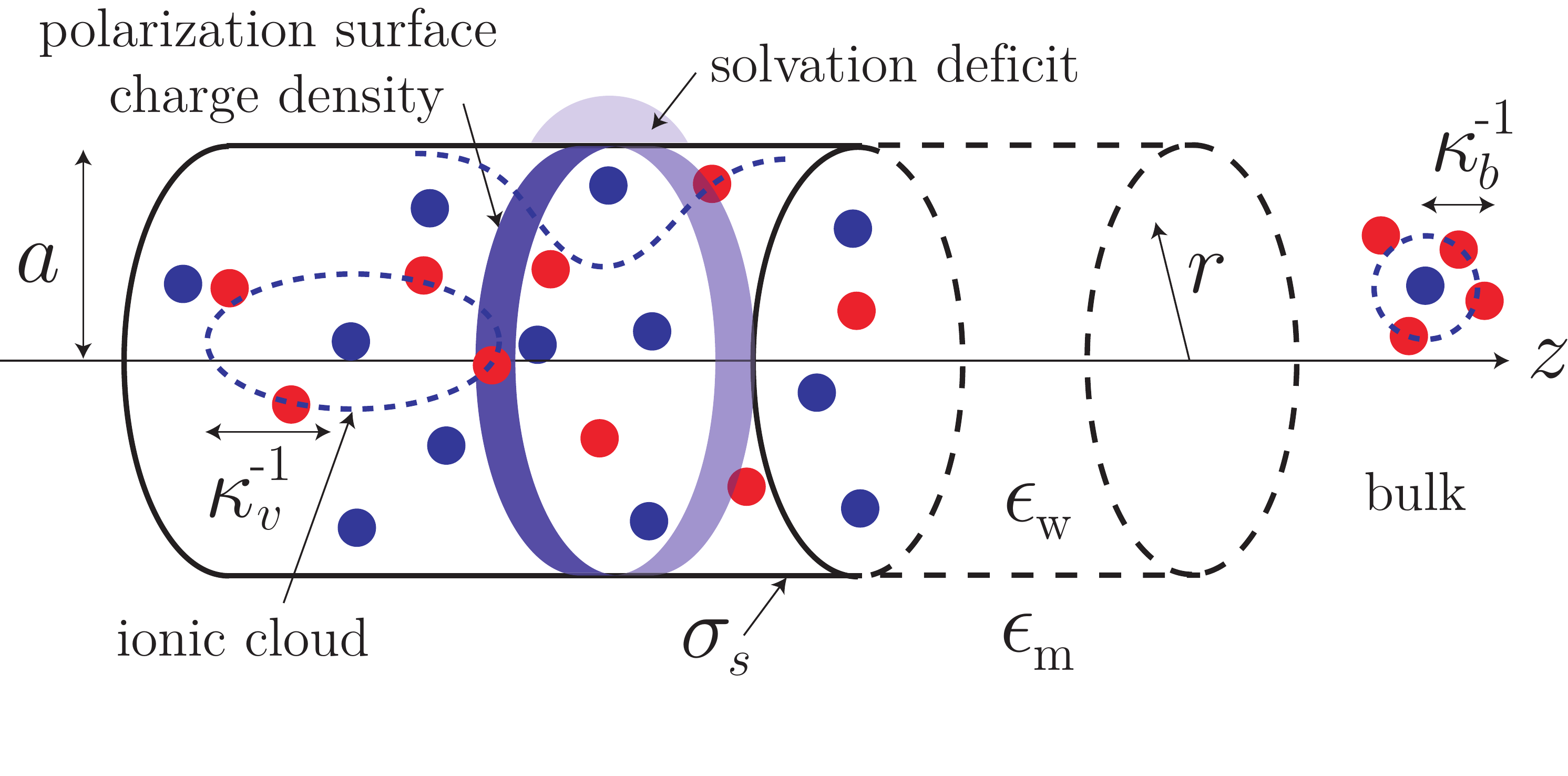}
\caption{Sketch of a cylindrical nanopore (radius $a$) filled with counterions and coions. The surface charge density is $\sigma_s$, $\kappa_v$ and $\kappa_b$ are screening parameters in the pore and in the bulk (cylindrical coordinates $r,z$).}
\label{fig1}
\end{figure}

In this paper, we show that when an electrolyte is confined to a neutral or weakly charged cylindrical nanopore, and in thermal and chemical equilibrium with an external salt solution reservoir, a novel type of ionic \emph{liquid-vapor} pseudo transition occurs for conventional electrolytes at room temperature, in contrast to the bulk, and within the experimental salt concentration, pore size, and pore wall surface charge density range.
{This pseudo phase transition (occuring in a \textit{quasi} one dimensional infinite system)} presents parallels with capillary evaporation of water in hydrophobic nanopores~\cite{evaporation}.
The driving mechanism is a competition between the {enhanced screening with ionic concentration of the dielectric repulsion and the increase of the surface tension associated with the deformation of the ionic cloud}, as sketched in Fig.~\ref{fig1} ({steric interactions do not seem to play an important role and therefore we adopt the point ion approximation}).
By using a field-theoretic variational approach, ionic correlations and polarization charge effects can be taken into account non-perturbatively~\cite{netz,Nous}. In the bulk, if $1/a_i$ is introduced as a cutoff in momentum space, this approach correctly predicts the \emph{existence} of a first-order ionic L-V transition~\cite{chineselevin}.
In \textit{neutral} nanopores, we also find a (pseudo) transition between a high concentration, conducting ionic liquid phase and a very low concentration insulating ionic-exclusion vapor phase; for \emph{weakly charged} nanopores, however, the pseudo transition is to a low conductivity counterion-only vapor phase, where coions are nearly entirely excluded from the pore and due to electroneutrality the counterion concentration is fixed almost entirely by the surface charge density.
We finally propose that the underlying mechanism controlling conductivity fluctuations as a function of $p$H and divalent ion concentration in certain artificial and biological nanopores~\cite{lev,bashford,bubbles} is a manifestation of the pseudo transition proposed here in a weakly charged nanopore.

We consider an electrolyte at {$T=300$~K} of dielectric permittivity $\epsilon_{\rm w}=78$, confined in a cylindrical nanopore of radius $a$, length $L$, and  surface charge density $\sigma_s$; the space outside the pore is salt free with a dielectric permittivity $\epsilon_{\rm m}<\epsilon_{\rm w}$ (Fig.~\ref{fig1}). The electrolyte is in contact with an external ion bulk reservoir at the end boundaries of the pore, which fixes the fugacity of ions inside the pore according to chemical equilibrium, $\lambda_i\equiv e^{\mu_i}/\Lambda^3=\lambda_{i,b}$, where $\mu_i$ is the chemical potential of ion $i=1\dots\alpha$ (energies are in units of the thermal energy $k_BT=1/\beta$) and $\Lambda$ is the De Broglie wavelength of an ion. Although ions interact through the bare Coulomb potential, $v^{\rm b}_{\rm C}(\br,\br')=\ell_B/|\br-\br'|$  in a bulk electrolyte, where $\ell_B=\beta e^2/(4\pi\epsilon_{\rm w})\approx 0.7$~nm, dielectric jumps at the nanopore boundaries yield a modified Coulomb potential $v_{\rm C}(\br,\br')$ obeying $-\nabla\epsilon(\br)\nabla v_{\rm C}(\br,\br')=\beta e^2\delta(\br-\br')$.
After performing a Hubbard-Stratonovich transformation and introducing a fluctuating field $\phi(\br)$, the grand-canonical partition function for ions in the nanopore can be written as $\mathcal{Q}=\int \mathcal{D}\phi\;e^{-H[\phi]}/\mathcal{Q}_{\rm C}$ where the Hamiltonian is~\cite{netz,Nous}
\be\label{H}
H[\phi]=\int \mathrm{d}\br\left[\frac{\epsilon(\br)}{2\beta e^2}  [\nabla\phi(\br)]^2-i\sigma(\br)\phi(\br)-
\sum_{i=1}^\alpha\tilde\lambda_i e^{i q_i \phi(\br)}\right],
\ee
$q_i$ is the ion  valency, $\mathcal{Q}_{\rm C}=-\frac12\mathrm{tr}\ln v_{\rm C}$, {$\sigma(\br)=\sigma_s\delta(r-a)$}, and $\tilde\lambda_i=\lambda_i e^{q_i^2v_{\rm C}^{\rm b}(0)/2}$. The electrostatic potential is given by $i\langle\phi(\br)\rangle$.

Evaluating the partition function using Eq.~(\ref{H}) is intractable due to non-linear terms, and we use the Gaussian variational method, which consists in computing the variational grand-potential $\Omega_v=\Omega_0+\langle H-H_0\rangle_0$, where the expectation value is evaluated with a variational Gaussian Hamiltonian
\be\label{HVar}
H_0[\phi]=\frac12 \int_{\br,\br'}\left[\phi(\br)-i\phi_0(\br)\right]v^{-1}_0(\br,\br')\left[\phi(\br')-i\phi_0(\br')\right],
\ee
and $\Omega_0=-\frac12\mathrm{tr}\ln(v_0/v_{\rm C})$. We then {extremize} $\Omega_v$ with respect to the variational functions, namely the Green's function $v_0(\br,\br')$ and the electrostatic potential $\phi_0(\br)$. This yields two intractable coupled non-linear differential equations, similar in form to a generalized Poisson-Boltzmann equation for $\phi_0(\br)$ and a generalized Debye-H\"{u}ckel (DH) one for $v_0(\br,\br')$~\cite{netz,Nous}. To make progress we consider the restricted case of a \emph{constant} $\phi_0$ in the pore~\cite{yarosh} and take $v_0(\br,\br')$ as the solution of
$[-\nabla\epsilon(\br)\nabla+\epsilon(\br)\kappa^2(\br)]v_0(\br,\br')=\beta e^2\delta(\br-\br')$
with a screening parameter, $\kappa(\br)=\kappa_v\,\Theta(a-r)$~\cite{onsager,yarosh}. We are then left with two variational parameters, the effective Donnan potential, $\phi_0$, and the DH parameter in the pore $\kappa_v$~\cite{Nous}.

{The variational grand-potential becomes $\Omega_v= -p \pi a^2L +\gamma^{\rm s}_v 2\pi a L$, where $p=\sum_i\lambda_i e^{q_i^2\ell_B\kappa_v/2}-\kappa_v^3/(24\pi)$ is the pressure of a bulk electrolyte with screening parameter $\kappa_v$~\cite{mcquarrie} and  $\gamma^{\rm s}_v$  a surface contribution given by
\bea
\gamma^{\rm s}_v &=& \sigma_s\phi_0-\frac{a}2\sum_{i=1}^\alpha\lambda_i e^{q_i^2\ell_B\kappa_v/2}\lan e^{-q_i^2\delta v_0(r,r)/2-q_i\phi_0}-1\ran \nonumber\\
&+&\frac{a\kappa_v^2}{16\pi\ell_B}\int_0^1 \mathrm{d}\xi\lan\delta v_0(r,r;\kappa_v\sqrt\xi)-\delta v_0(r,r;\kappa_v)\ran
\label{gamma}
\eea
where $\lan \dots \ran$ is the average over the nanopore and $v_0(\br,\br')=\ell_Be^{-\kappa_v|\br'-\br|}/|\br'-\br|+\delta v_0(\br,\br')$ with}
\be
\delta v_0(r,r;\kappa_v)=\frac{4\ell_B}{\pi}\int_0^\infty \mathrm{d}k\sideset{}{'}\sum_{m\geq0}F_m(k;\kappa_v)I_m^2(\varkappa r)
\ee
where $\varkappa^2=k^2+\kappa_v^2$, $\sum'$ means that the term $m=0$ is divided by 2,
\be
F_m(k;\kappa_v)=\frac{\epsilon_{\rm w}\varkappa K_m(ka)K'_m(\varkappa a)-\epsilon_{\rm m}k K_m(\varkappa a)K'_m(ka)}
{\epsilon_{\rm m}k I_m(\varkappa a)K'_m(ka)-\epsilon_{\rm w}\varkappa K_m(ka)I'_m(\varkappa a)}
\ee
and $I_m$ and $K_m$ are modified Bessel functions~\cite{janco,long}. {The first term in Eq.~(\ref{gamma}) is the electrostatic energy of the surface charge, the second term is a depletion term and the last term is the cost of ionic cloud deformation.}

In the following, we consider low to moderate ion concentrations in the bulk reservoir such that the  physical minimum of $\Omega_v$ in the bulk (for $a\to\infty$) is given by the DH result~\cite{Nous,yarosh},
$\lambda_{i,b}=\rho_{i,b}\,e^{-q_i^2\kappa_b\ell_B/2}$ where $\kappa_b^2=4\pi \ell_B \sum_{i=1}^\alpha q_i^2\rho_{i,b}$.
\begin{figure*}[!t]
\includegraphics[width=\linewidth]{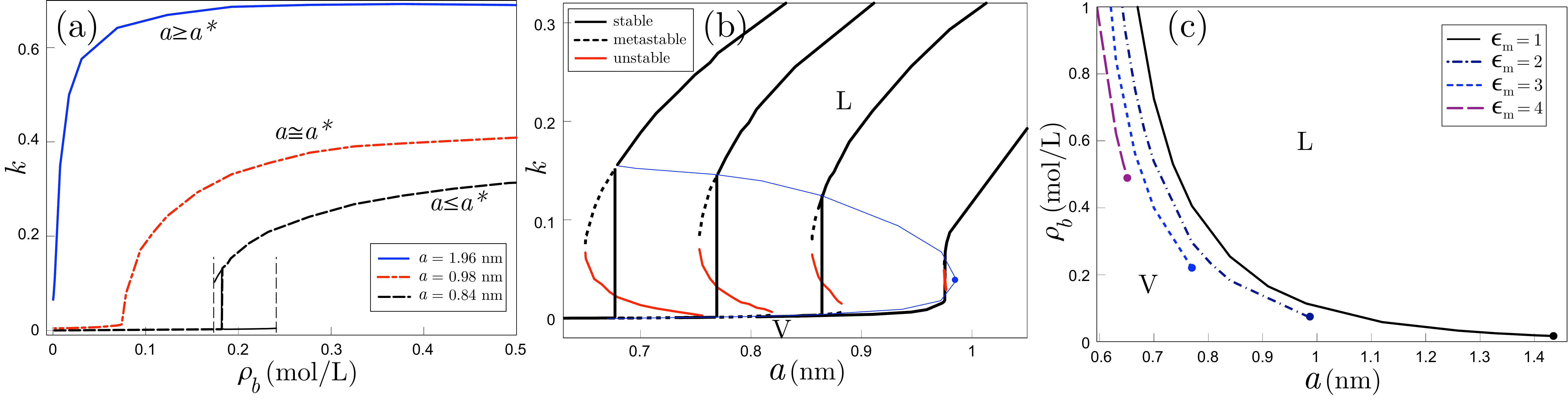}
\caption{Partition coefficient $k=\langle\rho\rangle/\rho_b$ inside a \textit{neutral} nanopore ($q=1$, $\epsilon_{\rm m}=2$, $\epsilon_{\rm w}=78$) vs (a)~the bulk concentration $\rho_b$ for three pore radii with metastable branches (thin solid lines) and window (vertical lines) for $a=0.84$~nm; and (b)~the pore radius $a$ for, from left to right, $\rho_b=0.7,0.3,0.156,0.08$~mol/L. Dotted (grey/red) lines show metastable (unstable) branches, light grey/blue lines (guide for the eye) are the ``boiling point'' curve (bottom) and the ``dew point'' curve (top) and the dot is the critical point, $\rho_b^*=0.074$~mol/L, $a^*=0.989$~nm. (c) Phase diagram for various $\epsilon_{\rm m}$ ($\epsilon_{\rm w}=78$). Critical lines correspond to phase boundaries between the ionic-penetration phase (L, above) and  the ionic-exclusion one (V, below).}
\label{fig2}
\end{figure*}
The ion concentration averaged over the pore section is $\langle\rho_i\rangle=\rho_{i,b} \Gamma_i e^{-q_i\phi_0}$ with $\Gamma_i\equiv \langle e^{-q_i^2w(r)/2}\rangle$ where the potential $w(r)$ incorporates the solvation energy due to the ionic cloud and polarization charge-ion interactions
\bea\label{PMF}
w(r)&=& (\kappa_b-\kappa_v)\ell_B+\delta v_0(r,r;\kappa_v)
\eea
and $\Phi_i(r)=q_i^2w(r)/2+q_i\phi_0$ is the potential of mean force of ion $i$ in the nanopore~\cite{Nous}.

For a \textit{symmetric} electrolyte ($q_\pm=q$, $\Gamma_\pm=\Gamma$), by extremizing $\Omega_v$ with respect to $\phi_0$, one obtains the electroneutrality condition, $\sigma_s=q\rho_ba\Gamma\sinh\left(q\phi_0\right)$, which, in the mean-field Poisson-Boltzmann limit ($w(r)=0$) leads to the usual Donnan potential~\cite{yarosh}. By injecting the solution for $\phi_0$ in $\Omega_v$, we are left with a single variational parameter $\kappa_v(\rho_b,a,\epsilon_{\rm m}/\epsilon_{\rm w})$. The averaged partition coefficients of counterions and coions are
\be
k_\pm\equiv\frac{\langle\rho_\pm\rangle}{\rho_b}=\Gamma e^{\mp q\phi_0}.
\label{partco}
\ee

In a \textit{neutral} nanopore, $\phi_0=0$ and $k_\pm=\Gamma$. For monovalent ions ($q=1$) and low enough {membrane} permittivities ($\epsilon_{\rm m}<5$), pore radii ($a<1$~nm) and bulk ion concentrations, the variational grand-potential $\Omega_v(\kappa_v)$ exhibits a minimum at $\kappa_v^{\rm V}a\simeq 5\times10^{-2}$. By increasing the reservoir concentration this minimum jumps discontinuously to a finite value $\kappa_v^{\rm L}$ with $\kappa_v^{\rm V}<\kappa_v^{\rm L}<\kappa_b$, and the pore undergoes a first-order pseudo phase transition from an \emph{ionic-exclusion} state to an \emph{ionic-penetration} one. As illustrated in Figs.~\ref{fig2}a and~\ref{fig2}b for $\epsilon_{\rm m}=2$, where $k$ is plotted, respectively, vs $\rho_b$ and $a$, for $a<a^*=0.989$~nm, a critical pore radius, a jump occurs for a specific {coexistence} value of $\rho_b$ larger than $\rho_b^*=0.074$~mol/L. At the transition, there is phase coexistence, $\Omega_v(\kappa_v^{\rm V})=\Omega_v(\kappa_v^{\rm L})$, where $\kappa_v^{\rm L}/\kappa_v^{\rm V}\simeq10$. When increasing $\rho_b$ beyond this coexistence value, the ionic-exclusion stable state becomes metastable and the pore becomes penetrable to ions (Fig.~\ref{fig2}a). For $a=a^*$, the pseudo transition is continuous.

Because of the strong ion depletion near the pore surface, an instructive analogy with capillary evaporation of water in hydrophobic nanopores favoring the vapor phase can be drawn~\cite{evaporation}. For ionic fluids, in contrast to water, we find that the critical temperature increases with decreasing pore radius.
Whereas the critical temperature $T_c$ of the L-V transition in bulk electrolytes is extremely low (reduced temperature $\tilde T^{\rm bulk}\equiv2 a_i/\ell_B\simeq0.05$ with $a_i\simeq 0.1$~nm  the ion radius), the phase diagram in Fig.~\ref{fig2}c shows that at room temperature, the exclusion pseudo phase transition takes place for experimentally accessible parameter values ($\epsilon_{\rm w}=78$, $\epsilon_{\rm m}=1$ to 4, and reduced temperature $\tilde T^{\rm pore}\equiv2 a/\ell_B\simeq1$): the coexistence lines separate the L-, above the line, from V-state, below the line. Beyond the critical point $(a^*(\epsilon_{\rm m}),\rho_b^*(\epsilon_{\rm m}))$, we enter the ``fluid'' phase where the pseudo transition disappears and is replaced by a smooth crossover (as in slit pores~\cite{Nous,Dresner,yarosh}). Using a simplified self-consistent approach, Dresner also found a first-order pseudo phase transition for electrolytes confined in neutral spherical pores~\cite{Dresner}, but Yaroschuk may have wrongly argued that this is an artefact arising from the use of the self-consistent DH equation~\cite{yarosh}.
When increasing $\epsilon_{\rm m}$ and thus decreasing the repulsive surface polarization charge, the parameter range where the phase separation is observable is considerably reduced.
{Hence, for neutral pores the ion penetration pseudotransition is driven by the competition between the last two terms of Eq.~(\ref{gamma}), which favor respectively high and low $\kappa_v$.}
\begin{figure*}[t]
\includegraphics[width=\linewidth]{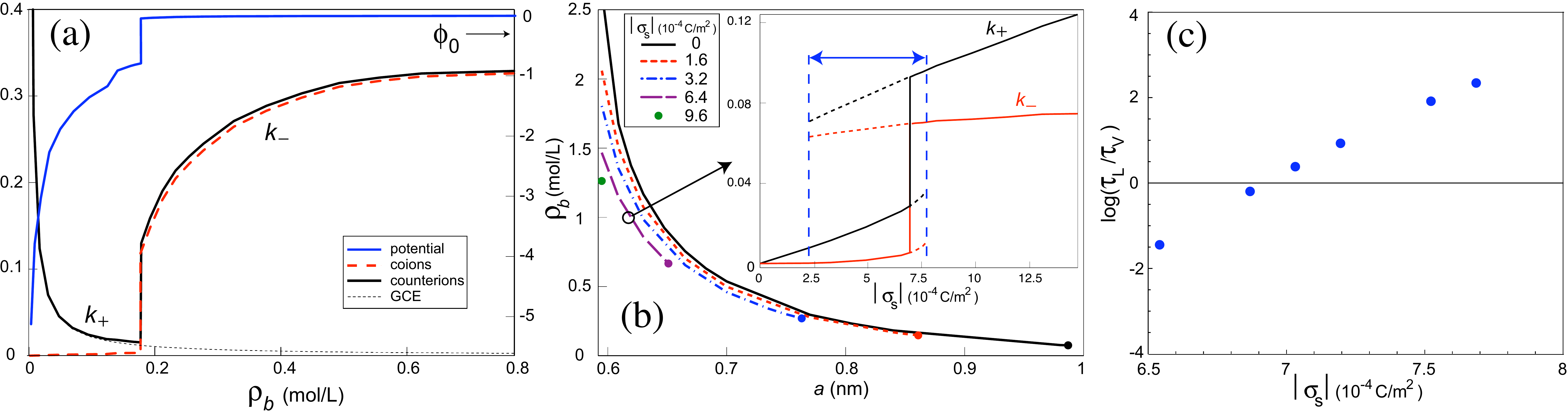}
\caption{(a) Partition coefficients of coions $k_-$ and counterions $k_+$ (left) and effective Donnan potential $\phi_0$ (right) vs. $\rho_b$   for charged pores ($\epsilon_{\rm m}=2$, $a=0.84$ nm, and fixed surface charge density $\sigma_s=-0.9\times10^{-4}$~C/m$^2$). The dotted line is the counterion-only approximation for counterions. (b) Phase diagram (similar to Fig.~\ref{fig2}c) for several values of $|\sigma_s|$. Inset: $k_\pm$ vs. $|\sigma_s|$ ($\rho_b=1$~mol/L, $a=0.617$~nm) showing the window (horizontal arrow) with stable (solid lines) and metastable branches (dashed lines); {the predicted HC to LC conductivity ratio $\propto \sum k_i^L/\sum k_i^V\simeq5$ is close to that seen in experiments~\cite{lev} (a value difficult to explain via water capillary evaporation~\cite{evaporation})}. (c) Logarithm of the ratio of resident times in L and V states of the inset of (b), $\tau_L/\tau_V=\exp[\Omega(\kappa_v^V)-\Omega(\kappa_v^L)]$ vs. $|\sigma_s|$ (pore length $L=5$~$\mu$m~\cite{lev}).}
\label{fig3}
\end{figure*}

What happens for \textit{slightly charged} nanopores? In Fig.~\ref{fig3}a are plotted the partition coefficients of counterions $k_+$ and coions $k_-$ and the variational Donnan potential $\phi_0$ vs $\rho_b$ for $\epsilon_{\rm m}=2$, $a=0.84$~nm and a weak surface charge, $\sigma_s=-8.6\times10^{-5}$~C/m$^2$. One observes that the discontinuous pseudo phase transition survives and phase coexistence occurs at $\rho_b=0.17$~mol/L, a value slightly lower than for the neutral case ($0.18$~mol/L). At this coexistence value, $\phi_0$ also exhibits  a jump, and for larger $\rho_b$, $\phi_0\simeq 0$ and we recover the neutral case. A slight difference between $k_+$ and $k_-$ remains due to electroneutrality, rewritten using Eq.~(\ref{partco}) as $k_+-k_-=2|\sigma_s|/(q\rho_ba)$.
For smaller $\rho_b$, $k_-\ll k_+$ while $k_+$ rapidly increases. We thus reach a low surface charge density counterion-only regime, with $ k_+\simeq2|\sigma_s|/q\rho_ba$ (Fig.~\ref{fig3}a, dotted line) and $k_-= \Gamma^2/k_+$~\cite{Nous}. In this regime, $\langle\rho_+\rangle\simeq2|\sigma_s|/(qa)$, is independent of $\rho_b$ and determined solely by global electroneutrality, and coions are excluded mainly by dielectric repulsion and not charge. Hence the vapor phase is no longer an ionic-exclusion phase but a \emph{weak ionic-penetration} one, governed by the weak surface charge density. For larger surface charge densities, $|\sigma_s|>10^{-3}$~C/m$^2$, the transition disappears.
The phase diagram for several values of $\sigma_s$ is illustrated in Fig.~\ref{fig3}b. Increasing the surface charge favors the L phase by reducing the coexistence line and shifting the critical point towards smaller $a$ and larger $\rho_b$ (because $\sigma_s$ increases $\gamma^{\rm s}_v$, but due to screening, this enhancement of $\gamma^{\rm s}_v$, decreases with $\kappa_v$). Comparison of Figs.~\ref{fig2}c and~\ref{fig3}b clearly shows that an increase in $|\sigma_s|$ plays qualitatively the same role as an increase in $\epsilon_{\rm m}$, i.e. a decrease in the repulsive polarization charge density.

Can the {trace of this first-order pseudo transition be observed experimentally in finite sized open-ended pores}? By studying ionic conductivity in nanopores ($a\simeq1$~nm, $L\simeq5~\mu$m) produced in polyethylene terephthalate membranes~\cite{lev}, Lev \textit{et al.} observed three regimes: a high conductivity (HC) regime,  a low conductivity (LC) one, and a two-state high/low conductivity (HC/LC) regime where the conductivity switches rapidly between both states. Interestingly, the dynamic characterization of the HC/LC regime, performed by measuring the ratio of resident times, leads to the identification of three parts: HC more stable, HC and LC at coexistence, and LC more stable. We argue that one possible mechanism for these current fluctuations {on the scale of seconds}  is that the system is close to the phase coexistence presented here, where the HC and LC regimes are identified, respectively, with the ionic-penetration ``liquid'' and ionic-weak penetration ``vapor'' phases, and the HC/LC regime with the window of $\sigma_s$ for which metastable branches exist (inset of Fig.~\ref{fig3}b).
{For finite open-ended nanopores, we expect the L-V pseudo transition to be rounded, with phase separation and hysteresis replaced by two-state fluctuations between pseudo-stable or metastable phases~\cite{evaporation}.}
For surfaces carrying carboxylic acid groups, $|\sigma_s|$ is an increasing function of $p$H~\cite{hijnen} (for more details see~\cite{long}), and Lev \textit{et al.}~\cite{lev} showed that the HC/LC regime only exists within a narrow $p$H window. At high $p$H, $|\sigma_s|$ is high and only the HC state exists; at low $p$H, $|\sigma_s|$ is low and only the LC state exists. Within this ``fluctuation'' $p$H window, using {the two-state approximation~\cite{evaporation}}, our model yields trends for ratio of resident times vs. $|\sigma_s|$ (Fig.~\ref{fig3}c) in qualitative agreement with the experimental ones vs. $p$H (Fig.~ 4 of~\cite{lev}). Similarly, increasing the concentration of trace divalent cations decreases the bare negative surface charge~\cite{healy} and the nanopore can likewise switch from the HC to the LC state~\cite{lev}.
Observations revealing a strong correlation between $\sigma_s$ and conductivity fluctuations~\cite{bashford} can also be explained via the charge-regulation mechanism~\cite{hijnen}, {because $\sigma_s$ depends} in a self-consistent way on local solution characteristics~\cite{long}.
Although other mechanisms have been proposed to explain nanopore conductivity fluctuations (gas bubbles~\cite{bubbles} or salt occlusions~\cite{siwy}), we believe that the one proposed here provides a natural explanation for a host of experimental trends~\cite{lev,bashford} and deserves further detailed investigation~\cite{long}. {One open question concerns the role of ion-pairing; such effects appears naturally in a 2nd order variational approach and are currently under study. bUsing the Fisher and Levin~\cite{bulk} bulk approach as a guideline, we expect higher order corrections to lead to quantitative, but not qualitative, changes to our results. To corroborate our predictions it would also be interesting to perform simulations, although it may be extremely difficult to properly include dielectric discontinuities and reach sufficiently long time scales ($\sim 1$~s)}.
\acknowledgments{We would like to thank B. Coasne for helpful discussions. This work was supported in part by the French ANR (project SIMONANOMEM No. ANR-07-NANO-055).}

\end{document}